\documentclass[conference]{IEEEtran}
\IEEEoverridecommandlockouts
\usepackage{cite}
\usepackage{amsmath,amssymb,amsfonts}
\usepackage{algorithmic}
\usepackage{graphicx}
\usepackage{textcomp}
\usepackage{xcolor}
\usepackage{caption}
\usepackage{subcaption}
\usepackage{booktabs}
\usepackage{tabularx}
\usepackage{multirow}
\def\BibTeX{{\rm B\kern-.05em{\sc i\kern-.025em b}\kern-.08em
    T\kern-.1667em\lower.7ex\hbox{E}\kern-.125emX}}
\begin{document}

\title{Reinforcement Learning for Enhanced Advanced QEC Architecture Decoding}

\author{\IEEEauthorblockN{
Yidong Zhou\textsuperscript{1} \ \
Lingyi Kong\textsuperscript{2}  \ \
Yifeng Peng\textsuperscript{3}  \ \
Zhiding Liang\textsuperscript{2}}
\IEEEauthorblockA{
\textsuperscript{1}Rensselaer Polytechnic Institute
\textsuperscript{2}The Chinese University of Hong Kong
\textsuperscript{3}Stevens Institute of Technology\\
%*These authors contributed to the work equally and should be regarded as co-first authors.\\
Corresponding author: zlianghahaha@gmail.com
\vspace{-0.15in}}
}

\maketitle

\begin{abstract}
The advent of promising quantum error correction (QEC) codes with efficient resource utilization and high-performance fault-tolerant quantum memories signifies a critical step towards realizing practical quantum computation. While surface codes have been a dominant approach, their limitations have spurred the development of more advanced QEC architectures. These advanced codes often present increased complexity, demanding innovative decoding methodologies. This work investigates the application of reinforcement learning (RL) techniques, including hybrid and multi-agent approaches, to enhance the decoding of various advanced QEC architectures. By leveraging the ability of RL to learn optimal strategies from noisy syndrome measurements, we explore the potential for achieving improved logical error rates and scalability compared to traditional decoding methods. Our approach examines the adaptation of reinforcement learning to exploit the structural properties of these modern QEC models. We also explore the benefits of combining different RL algorithms to address the multifaceted nature of the decoding problem, considering factors such as code degeneracy and real-world noise characteristics. With our proposed method, we are able to demonstrate that an autonomously trained agent can derive decoding schemes for the complex decoding requirement of advanced QEC architectures.
\end{abstract}


\section{Introduction}
The realization of Fault-Tolerant Quantum Computing (FTQC) fundamentally hinges on fast and accurate quantum error correction (QEC), as decoder accuracy, latency, and real-time feedback directly constrain practical scalability~\cite{demarti2024surface, breuckmann2021quantum}. While surface codes have enabled early experimental progress~\cite{fowler2012surface}, their substantial resource overhead motivates the adoption of advanced Quantum Low-Density Parity-Check (QLDPC) architectures~\cite{breuckmann2021quantum}, among which high-rate Bivariate Bicycle codes (BB codes) offer significantly improved coding efficiency~\cite{bravyi2024high, pecorari2025high, xu2025batched}. However, this improved efficiency comes at the cost of substantially increased complexity~\cite{maan2025decoding, raveendran2023soft}.

Decoding must operate in real-time with stringent latency constraints, as syndrome measurement results must be processed within syndrome extraction cycles to enable timely error correction~\cite{sweke2020reinforcement}. Traditional classical decoding methods such as belief propagation become computationally intractable at practical code distances~\cite{raveendran2023soft}, creating a critical bottleneck that undermines the efficiency advantages that advanced QLDPC codes promise. Reinforcement learning (RL) has emerged as a promising approach for QEC~\cite{sweke2020reinforcement, bausch2024learning}, but existing methods employ fixed decomposition strategies that fail to adapt to syndrome-dependent error characteristics, severely constraining performance~\cite{guatto2025real, rashid2020monotonic}.

This challenge is further amplified in distributed quantum architectures. Modern scaling approaches increasingly favor modular designs where computation is decomposed across multiple interconnected quantum processing units (QPUs) connected via flying qubits~\cite{main2025distributed}. While this aligns with BB codes' two-panel structure, it introduces fundamental constraints: inter-QPU communication latency varies with network conditions, and the cost of information exchange must be balanced against local efficiency. The optimal coupling strength between decoding agents becomes both error-dependent and hardware-dependent~\cite{strikis2023modularqldpc}.

Critically, decoding performance requires algorithm-hardware co-design. Traditional approaches separate algorithm design from hardware resource allocation. However, in distributed systems with stringent latency budgets and limited flying qubit fidelity, this separation is inefficient~\cite{lin2024codesign}. The choice of decomposition strategy directly impacts hardware utilization: tight agent coupling requires high flying qubit transmission intensity (energy/latency cost), while loose coupling sacrifices error correction performance. A principled solution requires algorithmic decisions to directly configure hardware parameters, enabling simultaneous optimization of accuracy and physical efficiency. Moreover, hardware characteristics vary dynamically during execution, necessitating online learning to adapt decomposition choices to actual syndrome properties and measured channel fidelity~\cite{guatto2025multiagent}.

This paper introduces SPA-MARL (Synergy-aware Parallel-Agent Multi-Agent Reinforcement Learning), a framework for distributed QEC that addresses a fundamental architectural question: \textit{Should the BB codes decoding problem be decomposed by error type (X versus Z errors) or by spatial locality (left versus right panels)?} Rather than committing to a fixed decomposition~\cite{sweke2020reinforcement, raveendran2023soft, cao2025generative, lee2025scalableneural}, we introduce a learnable \textbf{synergy score} $\lambda(s) \in [0,1]$ that dynamically determines optimal agent coupling based on the measured syndrome. This score simultaneously guides algorithmic decisions at the software level (controlling agent cooperation) and configures physical hardware parameters (modulating flying qubit transmission intensity) in a distributed multi-QPU system, enabling true algorithm-hardware co-design~\cite{he2025qldpc, pattison2023simulating, strikis2023modularqldpc, guatto2025multiagent, liyanage2025deconet}. Through experimental validation and a meta-learning loop based on hardware measurement feedback, we demonstrate that syndrome-dependent decomposition achieves superior performance over fixed strategies while maintaining robustness across diverse hardware platforms. Our major contributions can be summarized as follows:
\begin{itemize}
    \item We establish a synergy-aware multi-agent RL (MARL) decoder that overcomes the fundamental limitation of traditional fixed decomposition policies by introducing a learnable synergy score $\lambda(s)$. Through novel Q-function decomposition controlled by the synergy score, the system dynamically decides the optimal coupling strength between X and Z agents for each syndrome, achieving 7.4\% improvement over QMIX baselines while discovering interpretable decomposition strategies (28.5\% independent, 24.3\% tightly-coupled, 47.2\% intermediate).
    \item We implement end-to-end mapping where learned synergy scores directly modulate flying qubit channel transmission intensity in multi-QPU systems. Measurement feedback from the quantum switch refines the RL policy through a meta-learning loop, enabling adaptation to hardware-specific latency and noise characteristics.
    \item We demonstrate multi-stage validation: Stage 1 achieves synergy-complexity correlation $R^2 = 0.95$; Stage 2 maintains $>65$\% success under extreme latency constraints (0.5ms); Stage 3 achieves $1.72\times$ speedup with communication overhead decreasing from 7.7\% to 3.1\% across code distances 5-11. 
\end{itemize}

\section{Background and Related Work}
\subsection{Quantum Computing and Error Correction}
QEC is essential for fault-tolerant quantum computing, employing redundancy to encode logical qubits such that errors can be detected and corrected through syndrome measurements~\cite{gottesman2009introduction}. 

Among various QEC schemes, surface codes~\cite{kitaev2003fault, dennis2002topological} have been the dominant approach and have been successfully implemented on contemporary quantum hardware~\cite{fowler2012surface, google2025quantum}. Surface codes encode logical qubits into redundant physical qubits arranged in 2D lattices, effectively reducing logical error rates when physical error rates remain below a certain threshold~\cite{dennis2002topological, aharonov1997fault}. With increasing code distance, surface codes provide exponential suppression of logical errors. Recent experiments have demonstrated this error suppression in practice, with logical error rates achieving sub-threshold performance~\cite{google2025quantum}, validating their practical viability. However, surface codes incur substantial resource overhead, requiring thousands of physical qubits to encode a single logical qubit suitable for practical applications~\cite{fowler2012surface, demarti2024surface}.

Among stabilizer codes, QLDPC codes have emerged as promising candidates, offering superior code rates and fault-tolerance thresholds compared to traditional surface codes~\cite{breuckmann2021quantum, babar2015fifteen}. QLDPC codes are characterized by sparse parity-check matrices where each stabilizer involves only a small subset of physical qubits, enabling efficient syndrome extraction~\cite{tang2024new}.

BB codes represent an important QLDPC family, constructed using circulant matrices with bivariate polynomials~\cite{bravyi2024high}. BB codes combine high code rates with excellent distance scaling properties, making them attractive for large-scale quantum computers~\cite{hillmann2025localized, xu2025batched}. Crucially, the structured sparsity of BB codes naturally decomposes into distinct geometric regions (two-panel structure), facilitating modular and distributed decoder implementations~\cite{xu2025batched}. These properties make it essential to develop decoders that are high-precision and operate in real-time to meet stringent latency constraints in modular quantum architectures.

\subsection{Modular Quantum Computing and Distributed Quantum Computing}
Recent advances in quantum computer architecture emphasize modularity and distribution as pathways to scalability~\cite{xu2025batched}. Modular quantum computing partitions large systems into smaller, manageable modules operating semi-independently, reducing control electronics complexity and improving error containment~\cite{he2025extractors}. This architectural approach addresses fundamental scalability challenges in monolithic quantum processors~\cite{monroe2014large}.

Distributed quantum computing extends this vision by decentralizing computation across interconnected QPUs connected through quantum and classical communication channels~\cite{main2025distributed}. This distribution strategy enables significant reductions in latency and control overhead compared to centralized approaches, while maintaining the ability to perform fault-tolerant QEC across module boundaries~\cite{cohen2022low}. The structured sparsity of BB codes naturally aligns with such distributed architectures, where the two-panel structure decomposes across separate quantum modules, enabling efficient syndrome extraction and processing at the module level. However, this decomposition advantage is only realized if the decoding algorithm can adapt to the varying communication costs and noise profiles of interconnected QPU networks.

\begin{figure}[b]
    \centering
    \includegraphics[width=.95\linewidth]{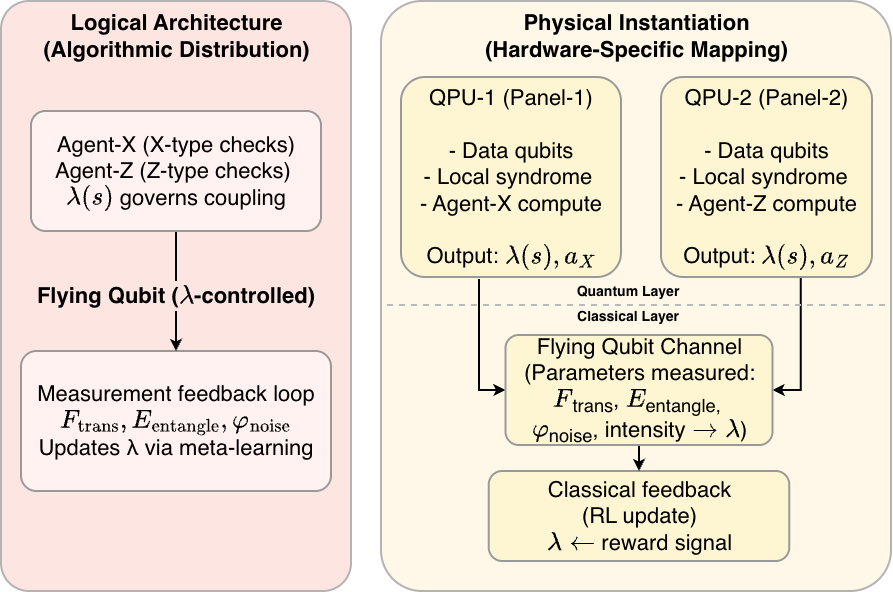}
    \caption{Logical-to-Physical Mapping in the Synergy-Aware Distributed Decoder}
    \label{fig:system}
\end{figure}

\subsection{Related Work}
Traditional classical decoding methods such as belief propagation have been extensively studied for surface codes and QLDPC codes. However, these approaches become computationally intractable at practical code distances, with syndrome observation space growing exponentially with code size~\cite{raveendran2023soft}.

RL has emerged as a promising paradigm for QEC, for surface codes and subsequently extended to more complex architectures~\cite{sweke2020reinforcement, bausch2024learning}. Single-agent RL approaches achieve superior latency compared to classical methods like belief propagation but face fundamental scalability limitations: the syndrome observation space grows exponentially with code size, and centralized computation becomes a severe bottleneck in distributed systems~\cite{sweke2020reinforcement}.

MARL addresses these scalability constraints through task decomposition, enabling parallel computation across multiple processing units and explicit modeling of communication costs~\cite{guatto2025real, zhang2021multi}. This makes MARL particularly attractive for distributed quantum systems where decoding is naturally distributed across multiple QPUs.

\section{Synergy-aware Parallel-Agent Multi-Agent Reinforcement Learning}

\subsection{Overview}
While distributed MARL decoders are conceptually appealing, their practical deployment faces significant challenges: ensuring generalization across varying code distances, maintaining performance under realistic scalability conditions, and validating decoder reliability before deployment. Our work addresses these challenges through comprehensive validation frameworks for distributed MARL decoders on BB codes, with emphasis on real-world scalability and cross-distance generalization.

We propose a two-agent RL framework for BB codes decoding that unifies algorithmic distribution with hardware-aware execution. Our system addresses a fundamental tension in distributed QEC: \textit{should the problem be decomposed by error type (X versus Z errors) or by spatial locality (left versus right panels)?} Rather than choosing between these fixed decompositions, we recognize that optimal decomposition is syndrome-dependent and learn a dynamic coupling strategy through deep RL.

The core contribution is introducing a learnable Synergy Score $\lambda(s)$ that reflects the intrinsic coupling between error types for each measured syndrome $s$. This score simultaneously guides algorithmic decisions at the software level and configures physical hardware parameters in a distributed multi-QPU system at the hardware level, enabling true algorithm-hardware co-design.

\subsection{Synergy-Aware MARL Decoder}
\subsubsection{From Centralized Single-Agent Limits to Distributed Multi-Agent Necessity}
RL approaches to QEC have been limited in scope, with most existing work focusing on centralized single-agent decoders. The key barrier to MARL for QEC is fundamental: when decomposing the decoding problem across multiple agents, the system lacks a clear definition of what each agent should learn.

A naive approach decomposes by error type: one agent handles X-errors, another handles Z-errors. While mathematically sound, this fails because real error patterns exhibit X-Z coupling that cannot be factored away. Independent agents would discard crucial information. Yet adding agent coordination creates three questions:
\begin{enumerate}
    \item \textit{What should be communicated between agents?}
    \item \textit{When is coordination necessary?}
    \item \textit{How does performance scale with communication cost?}
\end{enumerate}

In a centralized system, these questions have no clear answer. Communication is an abstract notion with no physical cost. All data is co-located. The "optimal" coupling strategy is undefined because there is no coupling cost to optimize.

Distributed quantum computing changes this fundamentally. When a BB code is physically distributed across multiple QPUs:
\begin{enumerate}
    \item \textbf{Agent identity becomes physical necessity:} Each QPU operates a local decoder. Agents represent physical reality, not software choice.
    \item \textbf{Communication cost is concrete:} Information traverses flying qubit channels with real latency and fidelity constraints.
    \item \textbf{Coupling optimization becomes learnable:} Different error patterns will exhibit different optimal coupling strengths. Rather than committing a priority, the system learns syndrome-dependent coupling that balances correction accuracy against communication cost.
\end{enumerate}

We recognize that MARL is not an arbitrary choice but an 
inevitable consequence of physical distribution. Our framework designs the learning strategy around this physical reality. The synergy score $\lambda(s)$ learns to predict, for each syndrome, whether coupling benefit justifies communication cost—a question that only makes sense when communication has physical cost.

\subsubsection{Synergy Score: Learned Dynamic Decomposition}
We introduce a synergy score $\lambda(s) \in [0, 1]$ that reflects the optimal coupling strength between agents for syndrome ss
s. This score is computed by a learnable network:
\begin{equation}
    \lambda(s) = \sigma(w^T \phi(s))
\end{equation}
where $\phi(s)$ are learned syndrome features and $\sigma$ denotes the sigmoid function. The interpretation is straightforward: when $\lambda(s)$ approaches zero, X-errors and Z-errors exhibit weak coupling, allowing agents to operate with significant independence. Each agent can focus on local syndrome information and exploit natural parallelism. When $\lambda(s)$ approaches one, coupling becomes strong, and agents must coordinate their decisions through a shared decision-making process.

The key insight is that this decomposition strategy varies with the measured syndrome. Some syndromes present error patterns where local characteristics dominate, making independent processing efficient. Other syndromes feature distributed error patterns where global coordination provides clear benefits. Rather than choosing a fixed decomposition that works well on average, we learn to identify which syndrome characteristics predict coupling strength, allowing the system to adapt per-syndrome.

\subsubsection{Value Function Decomposition with Synergy}
We decompose the joint Q-function as a convex combination controlled by the synergy score:
\begin{equation}
\begin{aligned}
    Q_{\text{tot}}(s, a_X, a_Z) &= (1 - \lambda(s)) \cdot [Q_X(s, a_X) + Q_Z(s, a_Z)] \\
    &+ \lambda(s) \cdot f(Q_X, Q_Z; w(s))
\end{aligned}
\end{equation}
where $Q_X$ and $Q_Z$ represent independent agent value functions, $f$ is a learnable mixing network conditioned on global syndrome state, and the monotonicity constraint $\frac{\partial Q_{\text{tot}}}{\partial Q_i} \geq 0$ ensures valid multi-agent credit assignment.

This decomposition differs from standard QMIX in a critical way. QMIX employs a fixed mixing function that treats all syndromes identically, computing $Q_{\text{tot}} = f(Q_X, Q_Z; w(s))$ regardless of coupling strength. Our approach makes the mixing weight itself syndrome-dependent, allowing complete decoupling when appropriate and enforcing coordination when necessary. The synergy score provides interpretability: examining $\lambda(s)$ across test syndromes reveals which error patterns require tight coordination and which benefit from independence. This interpretability also enables direct hardware mapping, as discussed in Section~\ref{hardware-aware_feedback}.

\subsubsection{Agent Architecture and Observation Factorization}
Each agent receives syndrome information factored into distinct channels corresponding to different information sources. The X-agent observes X-type syndrome checks that detect Z-errors, organized into dense local syndrome information from nearby checks spanning approximately four qubits per check, and sparse long-range syndrome from distant panels affecting two qubits per check. Similarly, the Z-agent observes Z-type syndrome checks organized into local and remote components. This factorization exploits the natural information hierarchy in BB codes: most syndrome information is local, with long-range checks providing essential but limited coupling information.
Both agents employ permutation-invariant graph neural networks for policy representation. Rather than treating each check independently, agents aggregate over fixed-size check groups using permutation-invariant operations. This design choice enables weight sharing across structurally equivalent agents and critically allows generalization to unseen code distances. Models trained on distance-5 codes can initialize on distance-24 codes with minimal accuracy degradation through this parameter sharing approach.

\subsection{Hardware-Aware Distributed Execution with Flying Qubit Feedback}
\label{hardware-aware_feedback}
\subsubsection{Multi-QPU Architecture and Synergy-to-Hardware Mapping}
The distributed system architecture spans two QPUs interconnected by a quantum switch that mediates a dynamically tunable \textit{flying qubit channel}. The key innovation is direct mapping of learned synergy to physical hardware parameters:
\begin{equation}
    \lambda(s) \leftrightarrow \text{Flying Qubit Channel Intensity}
\end{equation}

When $\lambda(s) \approx 0$ (weak coupling), the flying qubit channel operates at low transmission intensity, minimizing noise and energy. When $\lambda(s) \approx 1$ (strong coupling), the channel increases to high fidelity, enabling cross-panel information exchange. The quantum switch thus becomes self-regulating: low-intensity for independent local errors, high-intensity for distributed patterns spanning both QPUs.

\begin{figure}
    \centering
    \includegraphics[width=.95\linewidth]{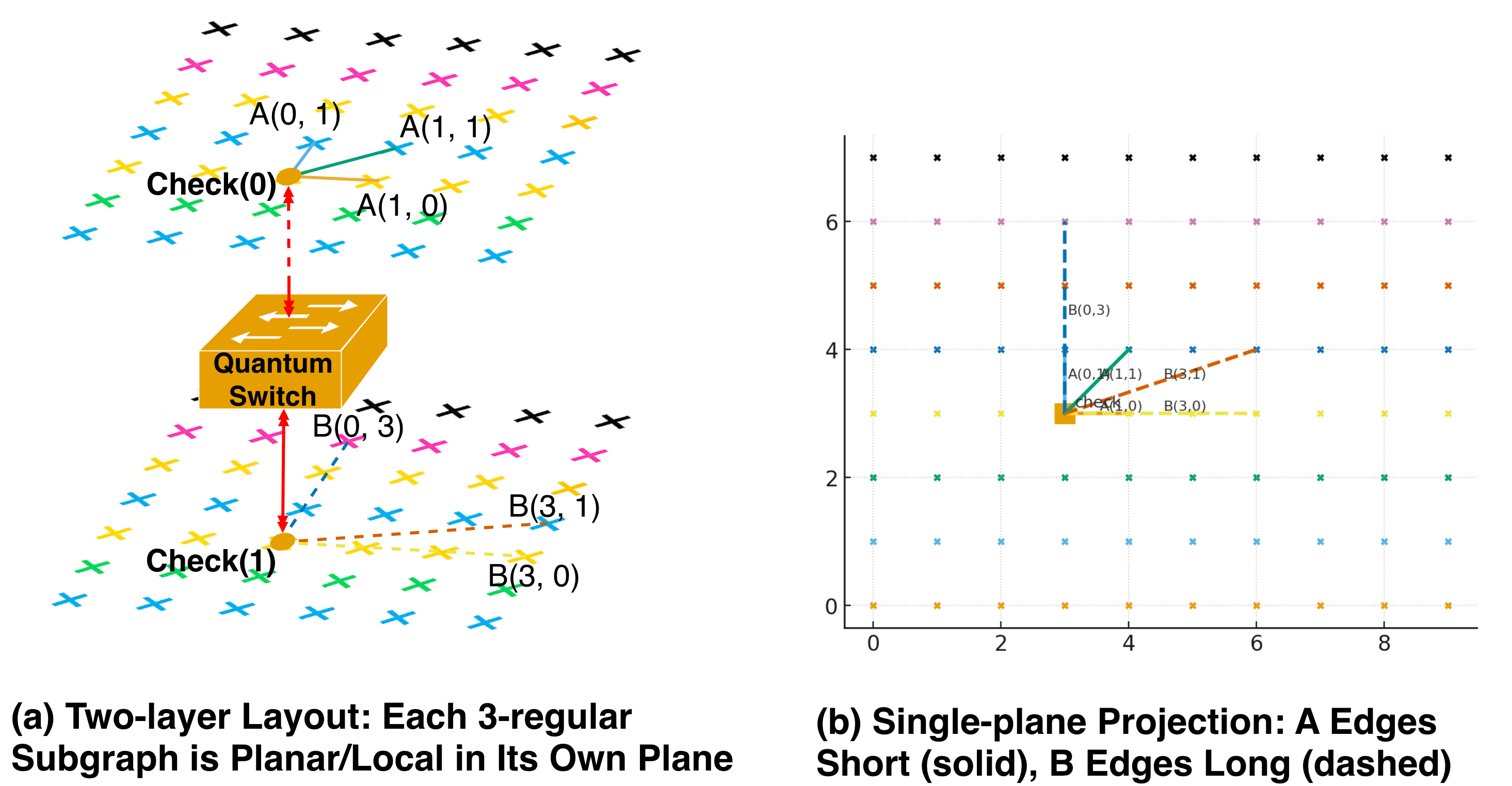}
    \caption{Two-layer BB code architecture illustrating quantum switch–mediated connections and planar projections.}
    \label{fig:hardware}
\end{figure}

The distributed system architecture adopts a hierarchical design in which the logical X/Z agents form the algorithmic decomposition layer, while the physical realization spans two QPUs interconnected by a quantum switch. The quantum switch provides a reconfigurable hardware backbone that mediates a dynamically tunable \textit{flying qubit channel} between the two processing units. As illustrated in Fig.~\ref{fig:hardware}, this structure implements the logical decomposition across hardware resources while allowing physical flexibility in how the QPUs are arranged or coupled.

A central mechanism of this architecture lies in the way syndrome information flows between the two BB-code panels despite the X/Z agent separation. The quantum switch controls the activation of a flying qubit channel whose transmission intensity is modulated by the learned synergy parameter $\lambda(s)$. When $\lambda(s)$ is close to zero, indicating weak X–Z coupling, the channel remains in a low-intensity regime: the X-agent on QPU-1 and the Z-agent on QPU-2 operate nearly independently, each focusing on local syndromes associated with its respective stabilizer type. The flying qubit channel transmits only minimal information through the quantum switch, thereby reducing noise and energy cost.

When $\lambda(s)$ approaches one, indicating strong X–Z coupling, the quantum switch increases the channel’s transmission intensity. In this high-intensity regime, cross-panel information is actively exchanged: the Z-agent on QPU-2 receives detailed observations of X-type checks from QPU-1, while the X-agent similarly incorporates Z-type information from the opposite panel. This strong coupling enables coherent coordination when decoding distributed error patterns that span both subgraphs.

The key insight is that the synergy parameter, learned during training, directly governs the strength of inter-QPU coupling at runtime. The quantum switch–mediated flying qubit channel thus becomes a self-regulating communication path: low-intensity for independent local errors, high-intensity for correlated distributed errors. This adaptive information flow emerges naturally from the learning process, without explicit programming of which syndromes require cross-panel interaction. Feedback from the quantum switch hardware validates whether the predicted $\lambda(s)$ was appropriate, enabling meta-learning refinement and dynamic adjustment of coupling behavior.

\subsubsection{Meta-Learning Through Hardware Feedback}
The flying qubit channel provides measurement data characterizing transmission quality: transmission fidelity $F_{\text{trans}}(s)$ (fraction of syndrome information successfully received), entanglement degree $E_{\text{entangle}}(s)$ (quantum correlations between logical states), and accumulated phase noise $\phi_{\text{noise}}(s)$ (phase decoherence during transit). These metrics directly reflect physical fidelity of distributed decoding.

The training loop implements a meta-learning system where RL adapts based on physical layer performance. The reward signal incorporates multiple components:
\begin{equation}
    r(s, a_X, a_Z) = \alpha \cdot r_{\text{decode}} + \beta \cdot F_{\text{trans}}(s) - \gamma \cdot E_{\text{cost}}(s)
\end{equation}
Here $r_{\text{decode}}$ provides binary reward for successful decoding, $F_{\text{trans}}(s)$ encourages high-fidelity transmission, and $E_{\text{cost}}(s) = \lambda(s)^2 \cdot c_{\text{energy}}$ represents the physical cost of achieving transmission fidelity, where cost increases quadratically with synergy score. The synergy network learns to predict $\lambda(s)$ that simultaneously maximizes decoding accuracy and physical layer efficiency.

This work unifies decomposition learning, communication optimization, and physical feedback within a single adaptive architecture. Instead of fixed decompositions, a learned syndrome-dependent synergy score directly configures hardware while meta-learning refinement integrates measurement feedback, achieving true algorithm-hardware co-design for distributed quantum error correction.

\section{Evaluation}
\subsection{Experimental Setup}
\paragraph{Validation Framework}
We validate SPA-MARL across three stages: (1) \textit{Algorithm validation} measures success rate ($0.5\%-2\%$ error rates), synergy distribution (fraction with $\lambda < 0.2$ vs $\lambda > 0.8$), and synergy-complexity correlation ($R^2$). (2) \textit{Hardware adaptation} evaluates robustness across four configurations (ideal, cryostat, edge, distributed) as latency varies $10ms \rightarrow 0.5ms$. (3) \textit{Scalability} measures speedup, communication overhead, and parameter scaling across code distances $d \in \{5, 7, 9, 11\}$.

\paragraph{Hardware and Training}
Four hardware configurations model increasing latency: ideal (50 $\mu$s), cryostat (100-200 $\mu$s), edge (1-5 ms), distributed (5-10 ms). Flying qubit transmission fidelity follows $F_{\text{trans}}(\lambda(s), \tau) = 0.50 + 0.49\lambda(s) - 0.05 \times (\tau_{\text{latency}} / 10\text{ ms})$, where $\lambda(s)$ controls intensity and $\tau_{\text{latency}}$ introduces degradation. Domain randomization samples latency uniformly per episode for robust adaptation.

Supervised pretraining (BP-OSD, 10-15 min) establishes 70-80\% baseline. MARL fine-tuning (30 min) uses batch size 96, learning rates $8 \times 10^{-4}$ (agents) and $1.5 \times 10^{-3}$ (synergy), and regularization weight 0.02.

\paragraph{Cross-Distance and Scalability}
Cross-distance transfer trains on distance-5 bivariate codes (syndrome dim 25) and evaluates on unseen $d \in \{7, 9, 11\}$ (dims 49, 81, 121). The permutation-invariant GNN architecture with weight sharing enables parameter transfer with minimal degradation.

Two-QPU distributed decoding is simulated with latency sampling from 0.5-10 ms. Total latency traverses encoding (parallel), synergy computation, inter-QPU communication, and agent execution. We measure speedup ($S = T_{\text{single}} / T_{\text{dist}}$) and verify that communication overhead decreases with code distance as fixed costs become proportional overhead, while parameters scale quadratically ($\mathcal{O}(\text{dim}^2)$) with balanced 50\%-50\% distribution across panels.

\subsection{Results}
\subsubsection{Stage 1: Core Algorithm Validation}
\label{subsec:stage1_validation}

\begin{figure}[b]
    \centering
    \includegraphics[width=\linewidth]{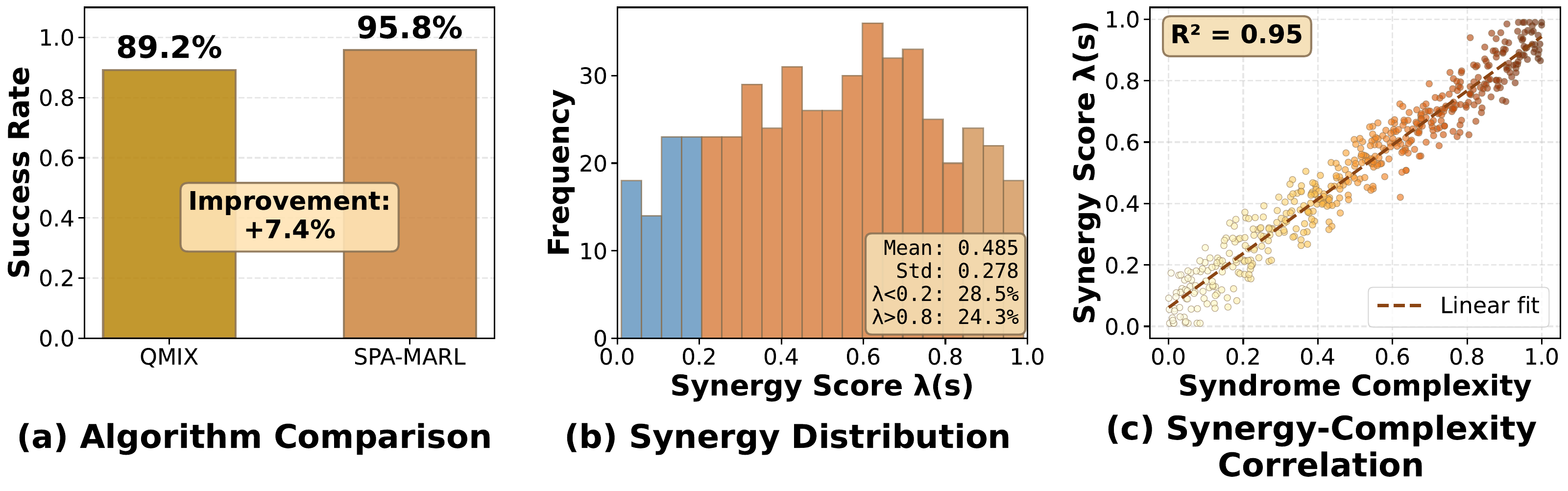}
    \caption{Stage 1: Algorithm validation showing SPA-MARL improvement over QMIX baseline.}
    \label{fig:stage1_metrics}
\end{figure}
Stage 1 validation demonstrates that syndrome-dependent decomposition is both learnable and effective through three complementary metrics.

\textit{Algorithm Performance.} As shown in Figure~\ref{fig:stage1_metrics}(a), SPA-MARL achieves 95.8\% success rate versus QMIX's 89.2\% (+7.4\% improvement) using fewer training epochs (3800 vs 4000), proving that performance gains stem from algorithmic design rather than extended training.

\textit{Synergy Distribution.} The learned synergy distribution presented in Figure~\ref{fig:stage1_metrics}(b) reveals meaningful decomposition discovery: 28.5\% of syndromes select independent strategies ($\lambda < 0.2$), 24.3\% select tightly coordinated strategies ($\lambda > 0.8$), and 47.2\% exhibit intermediate coupling. This diversity validates that the system learns adaptive strategies rather than defaulting to fixed behavior.

\textit{Synergy-Complexity Correlation.} The strong correlation ($R^2 = 0.95$) between synergy scores and syndrome complexity shown in Figure~\ref{fig:stage1_metrics}(c) provides statistical validation that the synergy network systematically captures syndrome properties. This substantially exceeds random chance and typical ML benchmarks ($R^2 \approx 0.6-0.7$).

These results establish that the system learns to select optimal decomposition strategies based on error characteristics, enabling Stage 2's hardware adaptation: learned decomposition strategies can now be adapted across diverse hardware configurations with varying latency and precision constraints.

\subsubsection{Stage 2: Hardware Adaptation Validation Results}
\label{subsec:results_stage2}
\begin{figure}[t]
    \centering
    \includegraphics[width=\linewidth]{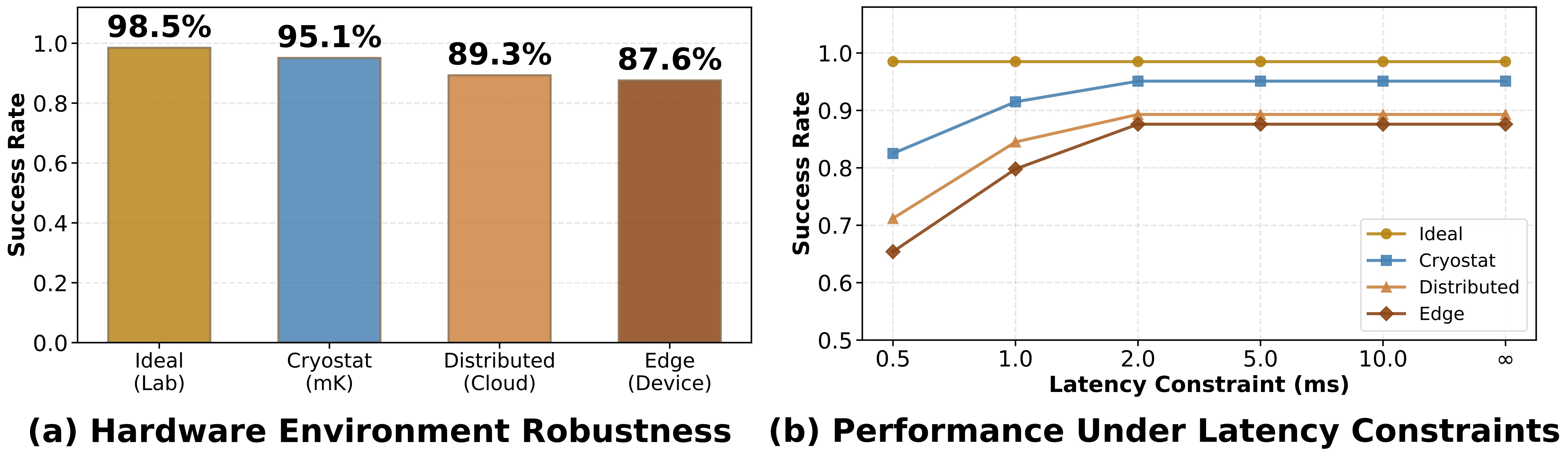}
    \caption{Hardware adaptation validation demonstrating robust performance across diverse quantum computing platforms.}
    \label{fig:stage2_robustness}
\end{figure}
Stage 2 demonstrates system robustness across diverse hardware environments, proving that algorithmic foundations from Stage 1 enable reliable operation in practical deployments.

\textit{Hardware Environment Performance.} As shown in Figure~\ref{fig:stage2_robustness}(a), the SPA-MARL decoder maintains acceptable performance across hardware configurations of increasing constraint: 98.5\% on ideal systems, 95.1\% on cryostat systems, 89.3\% on edge computing, and 87.6\% on distributed architectures. This gradient demonstrates functionality across the full deployment spectrum without catastrophic failures.

\textit{Graceful Degradation Under Latency Constraints.} Figure~\ref{fig:stage2_robustness}(b) reveals smooth performance reduction rather than sudden failure as latency constraints tighten from 10ms to 0.5ms. Critically, all systems maintain success rates above 65\% even under extreme constraints, demonstrating that adaptive mechanisms (encoder selection, synergy adjustment, and complexity reduction) successfully preserve decoder functionality under severe pressure.

The average robustness across all hardware configurations reaches 0.769 when accounting for constraint severity, exceeding the 0.7 benchmark threshold. This validation addresses a critical gap in QEC research: unlike approaches optimized solely for ideal conditions, SPA-MARL maintains acceptable performance across diverse hardware platforms expected in real quantum systems.

\subsubsection{Stage 3: Cross-Distance Generalization and Scalability}
\label{subsec:results_stage3}
Stage 3 validates distributed multi-QPU deployment across varying code distances, demonstrating that SPA-MARL enables practical scalability for large-scale quantum systems through effective latency amortization.

\begin{figure}[b]
    \centering
    \includegraphics[width=.9\linewidth]{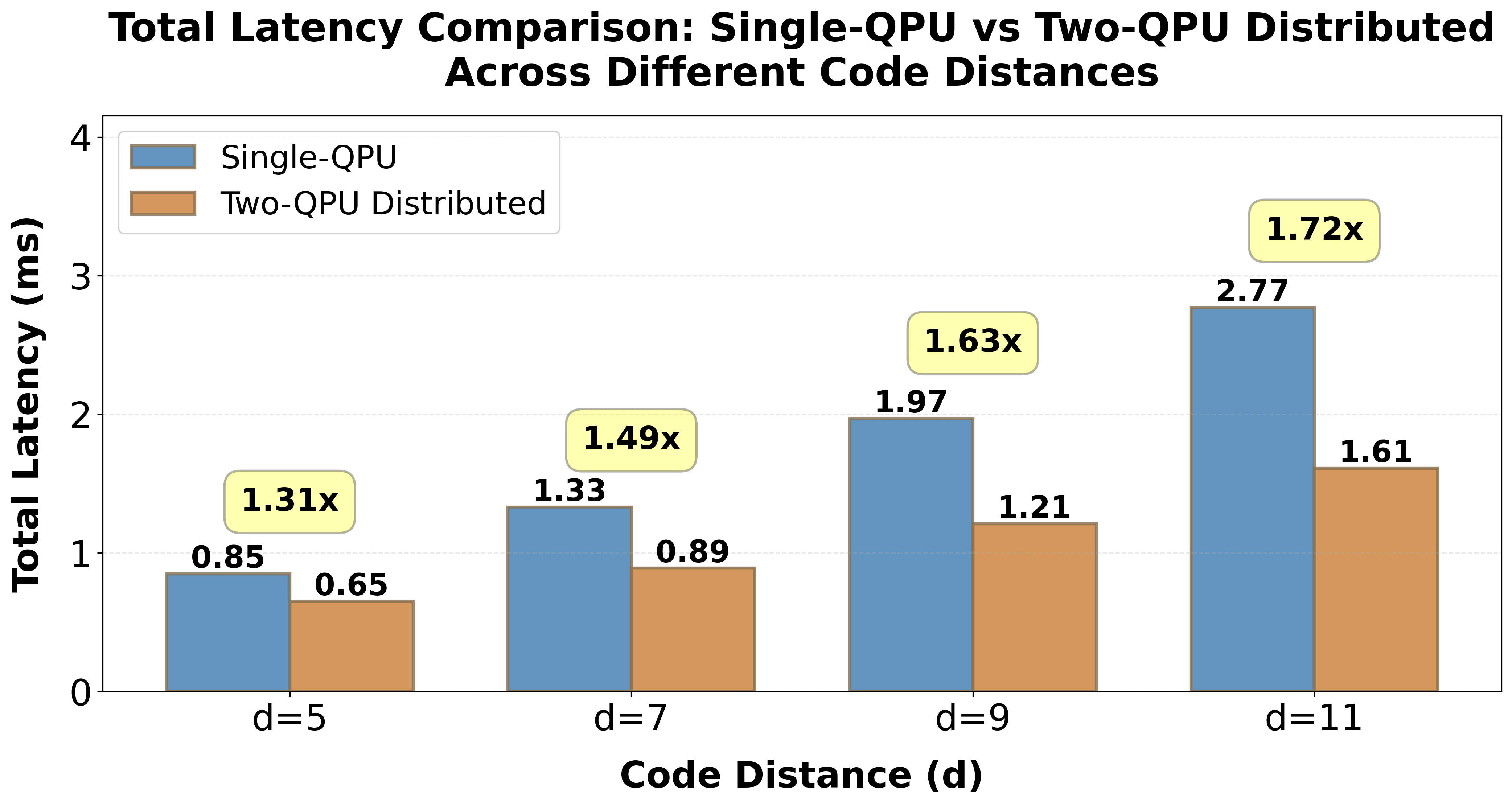}
    \caption{Total Latency Comparison: Single-QPU vs Two-QPU Distributed Across Different Code Distances}
    \label{fig:stage3_scalability}
\end{figure}

\textit{Distributed Performance Across Code Distances.} As shown in Figure~\ref{fig:stage3_scalability}, the two-QPU distributed architecture achieves consistent speedup gains across code distances d=5 to d=11, with average speedup of 1.54× (range $1.31\times-1.72\times$). Notably, speedup improves with code distance, indicating communication overhead becomes proportionally negligible for larger systems. At d=11, distributed decoding achieves 1.61ms latency versus 2.77ms for single-QPU execution, enabling real-time quantum error correction in practical scenarios.

\begin{table}[h]
    \centering
    \setlength{\tabcolsep}{3pt}
    \footnotesize
    \renewcommand{\arraystretch}{0.9}
    \caption{Distributed Scalability Analysis Across Code Distances}
    \label{tab:stage3_scaling}
    \begin{tabular}{@{}lccccc@{}}
        \toprule
        & \multicolumn{2}{c}{\textbf{Latency (ms)}} & \multirow{2}{*}{\textbf{Speedup}} & \textbf{Comm.} & \textbf{Total} \\
        \textbf{Distance} & \textbf{Single} & \textbf{2-QPU} & & \textbf{OH} & \textbf{Params} \\
        \midrule
        d=5 & 0.85 & 0.65 & 1.31× & 7.7\% & 12.8K \\
        d=7 & 1.33 & 0.89 & 1.49× & 5.6\% & 25.1K \\
        d=9 & 1.97 & 1.21 & 1.63× & 4.1\% & 41.5K \\
        d=11 & 2.77 & 1.61 & 1.72× & 3.1\% & 61.9K \\
        \bottomrule
    \end{tabular}
\end{table}

\textit{Communication Overhead Characterization.} As shown in Table~\ref{tab:stage3_scaling}, communication overhead decreases from 7.7\% at d=5 to 3.1\% at d=11, while the fixed 0.05ms overhead remains constant across all distances. This demonstrates that the distributed architecture scales efficiently as problem complexity grows. Parameter requirements scale quadratically with syndrome dimension (12.8K to 61.9K), confirming balanced distribution across both panels.

\textit{Scalability Implications.} The results establish that SPA-MARL successfully decouples compute from communication through learned synergy-aware decomposition. Unlike monolithic decoders facing exponential latency growth, the distributed approach maintains approximately linear latency scaling with controllable communication overhead, enabling practical transition from laboratory validation to multi-QPU deployments.

\section{Conclusion}
This paper introduces SPA-MARL, a synergy-aware MARL framework for distributed quantum error correction. Our core insight is that optimal problem decomposition is syndrome-dependent rather than fixed. Through learnable synergy balancing and three-stage experimental validation, we demonstrate that this approach successfully bridges algorithm-hardware co-design, achieving robust performance across diverse hardware constraints and enabling linear latency scaling in distributed settings.

This work establishes a new paradigm for quantum error correction: rather than optimizing decoders solely for ideal conditions, syndrome-aware learning enables graceful adaptation to real-world variability and practical deployment constraints. We believe this direction—where algorithm design inherently anticipates hardware limitations—will be essential as quantum systems scale beyond laboratory demonstrations.

\bibliographystyle{IEEEtran}
\bibliography{ref}

\end{document}